# Two objective and independent fracture parameters for interface cracks and a paradox

Jia-Min Zhao[1], He-Ling Wang[1], Bin Liu[1*]

[1] *AML, CNMM, Department of Engineering Mechanics, Tsinghua University, Beijing 100084, China*

*Corresponding author: Tel.: 86-10-62786194; fax: 86-10-62781824.

*E-mail address*: liubin@tsinghua.edu.cn

**Abstract**

Due to the oscillatory singular stress field around a crack tip, interface fracture has some peculiar features. This paper is focused on two of them. One can be reflected by a proposed paradox that geometrically similar structures with interface cracks under similar loadings may have different failure behaviors. The other one is that the existing fracture parameters of the oscillatory singular stress field, such as a complex stress intensity factor, exhibit some non-objectivity because their phase angle depend on an arbitrarily chosen length. Two objective and independent fracture parameters are proposed which can fully characterize the stress field near the crack tip. One parameter represents the stress intensity with classical unit of stress intensity factors. It is interesting to find that the loading mode can be characterized by a length as the other parameter, which can properly reflect the phase of the stress oscillation with respect to the distance to the crack tip. This is quite different from other crack tip fields in which the loading mode is usually expressed by a phase angle. The corresponding failure criterion for interface cracks does not include any arbitrarily chosen quantity, and therefore is convenient for comparing and accumulating experimental results, even existing ones.

**Introduction**

Since interface crack between dissimilar materials occurs in numerous applications and always leads to a failure, such as delaminating, many researchers have conducted investigations on both theoretical and experimental aspects. However, a comprehensive understanding on interface fracture still needs more efforts. The stress field around an interface crack tip was first pointed out with oscillatory singularities by Williams (1959), and then demonstrated by Erdogan (1963, 1965), Sih and Rice (1964), Rice and Sih (1965), England (1965), Suo (1989). The in-plane stress field near an interface crack tip between two different linear elastic isotropic materials (Fig. 1) can be expressed in polar coordinates $(r,\theta)$ as,

$$\sigma_{ij} = \frac{1}{\sqrt{2\pi r}} \left\{ \text{Re}\left(Kr^{i\varepsilon}\right) \Sigma_{ij}^{I}(\theta,\varepsilon) + \text{Im}\left(Kr^{i\varepsilon}\right) \Sigma_{ij}^{II}(\theta,\varepsilon) \right\} \tag{1}$$

where $\Sigma_{ij}^{I}(\theta,\varepsilon)$ and $\Sigma_{ij}^{II}(\theta,\varepsilon)$ are the angular functions and the corresponding analytical expressions are obtained by Rice et al. (1990) and given in Appendix A. $K$ is the complex stress intensity factor. The oscillatory parameter $\varepsilon$ is expressed as

$$\varepsilon = \frac{1}{2\pi} \ln\left(\frac{\mu_1 + \kappa_1 \mu_2}{\mu_2 + \kappa_2 \mu_1}\right) \tag{2}$$

where $\mu$ is shear moduli and $\kappa$ is a dimensionless parameter associated with Poisson's ratio $\nu$. Specifically, $\kappa = 3 - 4\nu$ for plane strain and $\kappa = (3-\nu)/(1+\nu)$ for plane stress. The subscript 1 and 2 mean the upper and lower materials in Fig.1, respectively. The oscillatory singularity of stress field around the crack tip brings many peculiar features, and some of them have been discussed and understood. For example,

**Peculiar feature #1 of interface crack: the crack surfaces overlap in the vicinity of the crack tip.**

The interpenetration of the crack surfaces near the crack tip was found theoretically by England (1965) and had confused many researchers for a long time. Many attempts have been made to eliminate this confusion, such as the contact zone model assumed by Comninou (Comninou, 1977, 1978; Comninou and Schmueser, 1979), the finite thickness interface model proposed by Atkinson (1977), nonlinear elastic behavior considered by Knowles and Sternberg (1983). Rice (1988) finally pointed out that this issue can be ignored because the dimension of the contact zone is not larger than an atomic bond length according to his estimation. Since then, the confusion on this peculiar feature has been almost addressed.

However, there are still some other peculiar features of interface cracks which have not been fully understood. Rice (1988) had demonstrated that, if two infinite bodies with interface cracks of different lengths subject to remote loading $\sigma_{yy}^{\infty}$ and $\sigma_{xy}^{\infty}$, as shown in Fig. 2, possess the same complex stress intensity factors $K$, the remote loading phase angle $\phi = \tan^{-1}\left(\sigma_{xy}^{\infty}/\sigma_{yy}^{\infty}\right)$ is different. In another word, if the loading phase angle are the same, i.e. $\phi_A = \phi_B$, the phase angles of $K$ for the different-sized interface cracks are different, which is against the instinct that the similar structures with different dimension under similar loadings should have the similar stress distributions. In our opinion, this contradiction, or a paradox, may lead to two questions:

(1) Is there any special fracture feature for similar structures with interface cracks under similar loadings?

(2) Is the complex stress intensity factors $K$ a proper parameter to characterize the interface crack field? If not, which one is better?

The second question has been discussed by some researchers, more detailed summary of the existing works will be given later. In this study, we will attempt to investigate these two questions. The paper is structured as follows. In Section 2, we propose another more straightforward paradox on geometrically similar structures with interface cracks to discuss question (1). Two objective and independent fracture parameters for the strange oscillatory singular stress field are derived in Section 3 to address question (2). How to apply the corresponding fracture criterion and measure the fracture toughness is discussed in Section 4. The main conclusions are summarized in Section 5.

## 2. A paradox on the failure of similar structures with interface cracks

As mentioned in the introduction section, the example or paradox discussed by Rice (1988) might include two influence factors. In this section, we only investigate the failure of geometrically similar structures with interface cracks, but do not adopt any fracture parameters, such as the complex stress intensity factor $K$. To make the contradiction prominent, a more straightforward paradox is designed as follows.

Figure 3 shows two identical specimens (Specimen A and B) with interface cracks subject to different loadings, i.e. tension and shear, respectively. When the loadings increase, the failure will happen. Obviously, these two specimens have different stress distributions near the crack tips and may have different failure modes (e.g., delamination along the interface or kinking) at their receptive critical failure moment. This difference is easy to understand due to the difference of external loadings. We then investigate two smaller specimens with similar geometries and external loadings to demonstrate an interesting contradiction.

According to the asymptotic stress field of an interface crack Eq. (1), especially the term $r^{i\varepsilon}$, the angular stress distribution varies periodically with the distance to the crack tip $r$, and is a linear combination of $\Sigma_{ij}^{I}(\theta,\varepsilon)$ and $\Sigma_{ij}^{II}(\theta,\varepsilon)$. As schematically shown in Fig. 3, at the rings denoted by the blue solid lines and red dashed lines, the angular distributions are pure $\Sigma_{ij}^{I}(\theta,\varepsilon)$ and $\Sigma_{ij}^{II}(\theta,\varepsilon)$, respectively. At the rings between the blue one and the red one, the angular distribution is a linear combination of $\Sigma_{ij}^{I}(\theta,\varepsilon)$ and $\Sigma_{ij}^{II}(\theta,\varepsilon)$. It can be imagined that for the identical specimens under different external loadings as Specimen A and B in Fig. 3, the pure angular distribution $\Sigma_{ij}^{I}(\theta,\varepsilon)$ always exists at a different distance to the crack tip, such as the radius $r_A$ for Specimen A and $r_B$ for Specimen B. We can then virtually extract two smaller specimens, Specimen A' with radius $r_A$ from Specimen A and Specimen B' with radius $r_B$ from Specimen B. The boundary loadings of these smaller specimens are the same as the stress when they are embedded in their corresponding original specimens, and have a pure angular distribution $\Sigma_{ij}^I(\theta,\varepsilon)$. Therefore, they have similar geometries and external loadings. It is interesting to note that as the subsystems of Specimen A and B, Specimen A' and B' have different stress distributions near the crack tips and may have different failure modes at their receptive critical failure moment. This is against our instinct and can be summarized as follows

**Peculiar feature #2 (or a paradox): Two geometrically similar structures with interface cracks under similar loadings may have different failure behaviors (e.g., failure modes and stress distributions near the crack tip).**

This feature cannot be observed for fracture in homogenous materials, and can be viewed as a special size effect that proportionally changing the dimension of a specimen with an interface crack results in significantly different failure behaviors. Moreover, it should be pointed out that the introduction of this paradox does not include any fracture parameters (e.g. $K$), and only the stress field is used.

The peculiar feature #2 (or the paradox) is so strange, because for any other solid systems two geometrically similar structures always have similar failure behaviors. We therefore doubt the correctness of this oscillatory singular stress filed for an interface crack. As we know, this stress solution has been derived analytically by many researchers (e.g., Williams, 1959; England, 1965; Erdogan, 1965; Rice and Sih, 1965). In the following, we will alternately use finite element method (FEM) simulations to check the solution. This is an almost impossible task since the oscillatory stress distribution only occurs within the region very close to the crack tip, which requires the finite element mesh size spanning over dozens orders and huge amount of degrees of freedoms. To overcome this difficulty, a recursive computing scheme is adopted in this paper. As schematically shown in Fig. 4, we repeat simulations on a finite element model with 86978 degree of freedoms (see Fig. 4(a)) under different angularly distributed displacement loadings. For each simulation, we extract the displacement distribution on the circle with the half radius, and apply it on the boundary of next simulation model. Repeating this process, a stress field of any circle can be obtained. More details can be found in Appendix B. It is found that the stress solution of an interface crack, Eq. (1) is validated. Therefore, the paradox (or peculiar feature #2) is still waiting for interpretations.

## 3 Independent and objective fracture parameters for an interface crack

### 3.1 Existing fracture parameters for interface cracks

How to choose proper fracture parameters to characterize the oscillatory singular stress field of an interface crack is another issue that has not been addressed completely or satisfactorily. In our opinion, proper fracture parameters should be objective quantities only depending on the stress field near the crack tip, and should be independent of other factors, such as size of sample, the length unit, or any length arbitrarily introduced by researchers. However, the existing fracture parameters are not objective in this sense.

*The complex stress intensity factor $K$*

The complex stress intensity factor $K$ has an awkward physical unit. Comninou (1990) pointed out that such a unit implies taking the logarithm of a length, which contradicts general ideas about units and dimensional consistency. Rice (1988) found that changing the length unit from meter to millimeter leads to the change of the phase angle of $K$. In particular, when the stress unit is $Pa$ and length unit is $l_{unit}$, $K$ can be expressed as

$$K = C \; Pa\sqrt{l_{unit}}\, l_{unit}^{-i\varepsilon} \tag{3}$$

where $C$ is a pure complex number.

According to Eq. (1), since the stress field near crack tip is objective, the term $Kr^{i\varepsilon}$ is an objective quantity. If two different length units $l_{unit}^{(1)}$ and $l_{unit}^{(2)}$ are adopted, we have

$$Kr^{i\varepsilon} = C^{(1)} Pa\sqrt{l_{unit}^{(1)}}\left(\frac{r}{l_{unit}^{(1)}}\right)^{i\varepsilon} = C^{(2)} Pa\sqrt{l_{unit}^{(2)}}\left(\frac{r}{l_{unit}^{(2)}}\right)^{i\varepsilon} \tag{4}$$

where $C^{(1)}$ and $C^{(2)}$ are corresponding pure complex numbers of the complex stress intensity factor $K$. $Pa\sqrt{l_{unit}^{(1)}}$ and $Pa\sqrt{l_{unit}^{(2)}}$ are pure real quantities. It is interesting to note that $C^{(1)}$ and $C^{(2)}$ must possess different phase angles because the phase angles of $\left(r/l_{unit}^{(1)}\right)^{-i\varepsilon}$ and $\left(r/l_{unit}^{(2)}\right)^{-i\varepsilon}$ are different, which again demonstrates that the phase angle of the complex intensity factor $K$ depends on the choice of length units. Therefore, $K$ is not an objective and proper fracture parameter.

*$K_I$ and $K_{II}$ with the unit of classical stress intensity factor*

As mentioned earlier in this section, $Kr^{i\varepsilon}$ is an objective quantity but $K$ is not.

One may introduce an arbitrarily chosen distance to the crack tip $\hat{r}$ and

$$Kr^{i\varepsilon} = K\hat{r}^{i\varepsilon} \cdot \left(\frac{r}{\hat{r}}\right)^{i\varepsilon} = \hat{K} \cdot \left(\frac{r}{\hat{r}}\right)^{i\varepsilon} = (K_I + iK_{II}) \cdot \left(\frac{r}{\hat{r}}\right)^{i\varepsilon} \tag{5}$$

where $\hat{K} = K_I + iK_{II} = K\hat{r}^{i\varepsilon}$ is defined by Rice (1988). Obviously, $K_I$ and $K_{II}$ have the unit of the classical stress intensity factor, and the influence of length unit is avoided. In the following, we demonstrate the dependence of $K_I$ and $K_{II}$ (or $\hat{K}$) on the choice of $\hat{r}$. When two different lengths $\hat{r}_A$ and $\hat{r}_B$ are adopted, according to Eq. (5) the corresponding $\hat{K}_A$ and $\hat{K}_B$ must satisfy

$$Kr^{i\varepsilon} = \hat{K}_A \cdot \left(\frac{r}{\hat{r}_A}\right)^{i\varepsilon} = \hat{K}_B \cdot \left(\frac{r}{\hat{r}_B}\right)^{i\varepsilon} \tag{6}$$

Thus,

$$\hat{K}_A = \hat{K}_B \cdot \left(\frac{\hat{r}_A}{\hat{r}_B}\right)^{i\varepsilon} \tag{7}$$

Based on the multiplication rule for complex numbers, it can be inferred that

$$\psi_A = \psi_B + \varepsilon \ln(\hat{r}_A/\hat{r}_B) \tag{8}$$

where $\psi_A$ and $\psi_B$ are the phase angles for $\hat{K}_A$ and $\hat{K}_B$, respectively, and $\varepsilon \ln(\hat{r}_A/\hat{r}_B)$ is the phase angle for complex number $(\hat{r}_A/\hat{r}_B)^{i\varepsilon}$. Equation (8) can also be found in existing literatures (Rice et al., 1990; Hutchinson and Suo, 1991) and theoretically proves that the phase angle of the complex intensity factor $\hat{K}$, i.e., $K_I$ and $K_{II}$, depends on the choice of arbitrarily chosen distance to the crack tip $\hat{r}$. Various characteristic lengths $\hat{r}$ have been suggested (Agrawal and Karlsson, 2007; Hutchinson and Suo, 1991), but $K_I$ and $K_{II}$ are not objective and proper fracture parameters in our opinion. Moreover, some researchers exhibited the measured critical failure loadings for $K_I$ and $K_{II}$ while $\hat{r}$ was also provided simultaneously (e.g., Ikeda et al., 1998), which is not convenient for comparison between different fracture tests if other $\hat{r}$ is taken.

Therefore, these imperfections of existing fracture parameters can be summarized as

**Peculiar feature #3 of interface crack: it is difficult to find objective and proper**

**fracture parameters to characterize the oscillatory singular stress field at a crack tip.**

To address the issue on fracture parameters, we adopt a new approach. Noting that $K_I$, $K_{II}$ and $\hat{r}$ can fully characterize the stress field at the tip but depend each other, we attempt to find two independent fracture parameters in the following.

**3.2 Two objective and independent fracture parameters for interface cracks**

Substituting Eq. (5) into Eq. (1) yields

$$\sigma_{ij} = \frac{1}{\sqrt{2\pi r}}\left\{\text{Re}\left[\hat{K}\left(\frac{r}{\hat{r}}\right)^{i\varepsilon}\right]\Sigma_{ij}^{I}(\theta,\varepsilon) + \text{Im}\left[\hat{K}\left(\frac{r}{\hat{r}}\right)^{i\varepsilon}\right]\Sigma_{ij}^{II}(\theta,\varepsilon)\right\} \tag{9}$$

Considering that

$$\begin{aligned}\hat{K}\left(\frac{r}{\hat{r}}\right)^{i\varepsilon} &= |\hat{K}|[\cos(\psi)+i\sin(\psi)]\left\{\cos\left[\varepsilon\ln\left(\frac{r}{\hat{r}}\right)\right]+i\sin\left[\varepsilon\ln\left(\frac{r}{\hat{r}}\right)\right]\right\} \\ &= |\hat{K}|\left\{\cos\left[\psi+\varepsilon\ln\left(\frac{r}{\hat{r}}\right)\right]+i\sin\left[\psi+\varepsilon\ln\left(\frac{r}{\hat{r}}\right)\right]\right\} \\ &= |\hat{K}|\left\{\cos\left[\varepsilon\ln\left(\frac{r}{\hat{r}e^{-\frac{\psi}{\varepsilon}}}\right)\right]+i\sin\left[\varepsilon\ln\left(\frac{r}{\hat{r}e^{-\frac{\psi}{\varepsilon}}}\right)\right]\right\}\end{aligned} \tag{10}$$

where $\psi = \arctan(K_{II}/K_I)$ is the phase angle of $\hat{K}$. Substituting Eq. 错误!未找到引用源。(10) into Eq. (1) yields

$$\sigma_{ij} = \frac{1}{\sqrt{2\pi r}}\left\{|\hat{K}|\cos\left[\varepsilon\ln\left(\frac{r}{\hat{r}e^{-\frac{\psi}{\varepsilon}}}\right)\right]\Sigma_{ij}^{I}(\theta,\varepsilon) + |\hat{K}|\sin\left[\varepsilon\ln\left(\frac{r}{\hat{r}e^{-\frac{\psi}{\varepsilon}}}\right)\right]\Sigma_{ij}^{II}(\theta,\varepsilon)\right\}. \tag{11}$$

By observing the equation above, we may introduce only two fracture parameters

$$K^{*} = |\hat{K}| = \sqrt{K_I^2 + K_{II}^2} \tag{12}$$

$$L_I = \hat{r}e^{-\frac{\psi}{\varepsilon}} \tag{13}$$

to fully characterize the stress field, and Eq. (1) becomes

$$\sigma_{ij} = \frac{K^{*}}{\sqrt{2\pi r}}\left\{\cos\left(\varepsilon\ln\frac{r}{L_I}\right)\Sigma_{ij}^{I}(\theta,\varepsilon) + \sin\left(\varepsilon\ln\frac{r}{L_I}\right)\Sigma_{ij}^{II}(\theta,\varepsilon)\right\}. \tag{14}$$

Although the proposed fracture parameters $K^{*}$ and $L_I$ are obtained from a specific $\hat{r}$, it can be proved that they are independent of $\hat{r}$. For the same stress field, two different lengths $\hat{r}_A$ and $\hat{r}_B$ correspond to the phase angle $\psi_A$, $\psi_B$ and the length

$L_I^A$, $L_I^B$. According to Eq. (8) and (13), we have

$$\frac{L_I^A}{L_I^B} = \frac{\hat{r}_A e^{-\frac{\psi_A}{\varepsilon}}}{\hat{r}_B e^{-\frac{\psi_B}{\varepsilon}}} = \frac{\hat{r}_A}{\hat{r}_B} e^{\frac{\psi_B - \psi_A}{\varepsilon}} = \frac{\hat{r}_A}{\hat{r}_B} e^{\ln\left(\frac{\hat{r}_B}{\hat{r}_A}\right)} = 1. \tag{15}$$

Therefore $L_I$ is independent of $\hat{r}$. By observing Eq. (14), $K^*$ is also independent of $\hat{r}$, so $K^*$ and $L_I$ are objective fracture parameters.

The physical meanings of $K^*$ and $L_I$ can be obtained from Eq. (14). $K^*$ represents the stress intensity and has classical unit of stress intensity factors. $L_I$ represents a characteristic length, and when $r = L_I$, the stress has pure angular distribution $\Sigma_{ij}^I(\theta, \varepsilon)$. Obviously, different loading modes lead to different $L_I$. We therefore name $L_I$ a loading mode length, which can properly reflect the phase of the stress oscillation with respect to the distance to the crack tip. By contrast, it is interesting to note that previous studies attempted to adopt a phase angle representing the loading mode but found its dependence on a chosen length. Moreover, $K^*$ and $L_I$ are independent of each other, and Eq. (14) express the stress field in a simple and easy to understand form. We recommend using this expression to avoid previous confusions.

It should be pointed out that if a given stress field is characterized by $K^*$ and $L_I$, $L_I' = e^{2\pi n/\varepsilon} L_I$ ($n$ is an integral) is also a correct loading mode length according to Eq. (14), which is similar to the phase angle in other problems both $\psi$ and $\psi + 2\pi$ are correct.

For an interface crack, the fracture criterion based on $K^*$ and $L_I$ can then be stated as,

$$\begin{cases} K^* \geq K_c^*(L_I), & \text{crack grows} \\ K^* < K_c^*(L_I), & \text{no crack growth} \end{cases} \tag{16}$$

where $K_c^*$ can be regarded as the fracture toughness under a loading mode with the length $L_I$.

# 4 Discussion

In this section, we will discuss how to apply the proposed $K^* - L_I$ based fracture criterion and measure the corresponding fracture toughness.

## 4.1 Converting existing experimental results for the proposed fracture criterion

Some researchers adopted fracture parameters $K_I$, $K_{II}$ and $\hat{r}$ to describe the failure behavior of interface cracks (Ryoji et al., 1994; Ikeda et al., 1998; Ji, 2016). According to Eqs. (12) and (13), the measured values of $K_I$, $K_{II}$ and $\hat{r}$ at the critical failure moment can be converted to $K_c^*$ and $L_I$ easily.

Moreover, some $K_I - K_{II} - \hat{r}$ based fracture criteria can be translated to $K^* - L_I$ based one. For example, Ikeda et al. (1998) obtained critical $K_I$ and $K_{II}$ of the aluminum-epoxy resin system for $\hat{r} = 0.01 mm$ in their experiments, and found that the $K_I - K_{II}$ failure curve can be represented by a partial ellipse, i.e.,

$$\left(\frac{K_I}{K_{Ic}}\right)^2 + \left(\frac{K_{II}}{K_{IIc}}\right)^2 = 1 \tag{17}$$

where $K_{Ic}$ and $K_{IIc}$ are the critical values for $K_I$ and $K_{II}$ at a given length $\hat{r}$, respectively. For different $\hat{r}$, $K_{Ic}$ and $K_{IIc}$ are different. Based on Eqs. (12) and (13), $K_I = K^* \cos\left[\varepsilon \ln\left(\hat{r}/L_I\right)\right]$ and $K_{II} = K^* \sin\left[\varepsilon \ln\left(\hat{r}/L_I\right)\right]$. Equation (17) can then be written in terms of $K^*$ and $L_I$ as

$$\left(\frac{K^*}{K_{Ic}}\right)^2 \cos^2\left(\varepsilon \ln \frac{\hat{r}}{L_I}\right) + \left(\frac{K^*}{K_{IIc}}\right)^2 \sin^2\left(\varepsilon \ln \frac{\hat{r}}{L_I}\right) = 1 \tag{18}$$

where $\hat{r}$, $K_{Ic}$ and $K_{IIc}$ have been provided in previous experiments. Therefore, this fracture criterion based on $K_I$, $K_{II}$ and $\hat{r}$ can be easily converted to the one based on $K^*$ and $L_I$.

Other researchers (Agrawal and Karlsson, 2007; Wang and Suo, 1990; Banks-Sills, 2015) suggested the following fracture criterion

$$G = G_c(\psi). \tag{19}$$

where $G$ is the energy release rate for an interface crack and $G_c$ is the fracture toughness which depends on the phase angle $\psi = \arctan(K_{II}/K_I)$. $G_c(\psi)$ is obtained by experiments for a given $\hat{r}$. The previous studies (Banks-Sills, 2015) provide the relation among $K_I$, $K_{II}$ and $G$ as

$$G = \frac{1}{H}\left(K_I^2 + K_{II}^2\right) \tag{20}$$

where $H = 2\cosh^2(\pi\varepsilon)/(1/\bar{E}_1 + 1/\bar{E}_2)$. $\bar{E} = 2\mu(1+v)$ for plane stress and $\bar{E} = 2\mu/(1-v)$ for plane strain. The subscript 1 and 2 denote the upper and lower materials in Fig.1 respectively. Substituting Eq.(12) (13) (20) into Eq. (19) yields

$$\frac{K^{*2}}{H} = G_c\left(\varepsilon \ln \frac{\hat{r}}{L_I}\right) \tag{21}$$

This then becomes a fracture criterion based on $K^*$ and $L_I$.

It should be emphasized that for the same bi-material but with different $\hat{r}$ chosen by different research groups, the unified $K^* - L_I$ based fracture criterion can be obtained, which is convenient for comparing and accumulating experimental results, even existing ones.

## 4.2 The size of $K^*$-dominant zone for interface cracks

When determining $K^*$ and $L_I$ in experiments, one must be aware of the size of $K^*$-dominant zone $r_{K^*}$. Within the region near the tip, $r \leq r_{K^*}$, the relative error between the full-field stress solution and asymptotic stress solution in Eq. (1) is smaller than a maximum acceptable error tolerance. We investigate the size of $K^*$-dominant zone of an infinite bi-material body including an interface crack subject to remote uniaxial tension as shown in Fig.5. The relative error between the full field solution and the asymptotic solution is defined in two ways as

$$\delta_{K^*}^{(1)} = \left|\frac{\sigma_{\theta\theta}^F\big|_{\theta=0} - \sigma_{\theta\theta}^A\big|_{\theta=0}}{\sigma_{\theta\theta}^F\big|_{\theta=0}}\right| \tag{22}$$

$$\delta_{K^*}^{(2)} = \frac{\sqrt{\int \left(\sigma_{\theta\theta}^F - \sigma_{\theta\theta}^A\right)^2 d\theta}}{\sqrt{\int \left(\sigma_{\theta\theta}^F\right)^2 d\theta}} \qquad (23)$$

where $\sigma_{\theta\theta}^F$ and $\sigma_{\theta\theta}^A$ are full field solution and asymptotic solution for hoop stress component, respectively. The first relative error has been used to evaluate the dimension of $K$-field (e.g., Anderson and Anderson, 2005), but we think the second error index represents the overall difference. The full field solution for this problem can be obtained from previous theoretical works (England, 1965; Erdogan, 1965; Rice and Sih, 1965) or finite element simulations (O'dowd et al., 1992; Becker et al., 1997). Figure 6 shows the relative errors for the cases with different stiffness ratios. It is found that if an error tolerance is given, the dimension of $K^*$-dominant zone is roughly the same as the one for $K$, but the predicted zone-sizes from two types of relative errors may differ several times.

A small $K^*$-dominant zone might bring difficulties in experimental measurements to determine $K^*$ via a direct matching approach. Two solutions can be considered alternatively to compute $K^*$ and $L_I$: one is applying the measured displacement of the outer region into FEM simulations as done in this paper; the other one is adopting some path-independent integrals, such as J-integral or M-integral (Chen and Shield, 1977; Yau et al., 1980; Yau and Wang, 1984; Bank-Sills et al., 1999).

## 5 Conclusion

In this paper, several peculiar features of interface crack are summarized and discussed. We mainly focus on finding objective and independent fracture parameters for the strange oscillatory singular stress field, and therefore superior to existing ones, such as a complex stress intensity factors dependent of length units. Specifically, two new fracture parameters $K^*$ and $L_I$ are proposed which can fully characterize the stress field near the crack tip. $K^*$ represents the stress intensity with classical unit of stress intensity factors, and $L_I$ is a characteristic length reflecting the loading mode. The previous studies have demonstrated that the loading mode cannot be characterized by a phase angle independent of a chosen length. The corresponding proposed failure criterion for interface cracks does not include any arbitrarily chosen quantity, and therefore is convenient for comparing and accumulating experimental results, even existing ones. Besides, another interesting peculiar feature or a paradox is identified that two geometrically similar structures with interface cracks under similar loadings may have different failure behaviors. This paradox can be partially interpreted with the fracture criterion including the loading mode length $L_l$. If a structure with interface crack under loadings is magnified in size proportionally, the length $L_l$ will increase accordingly, implying the change of the loading mode and failure behaviors. However, this paradox is still against our instinct and waiting for further interpretations.

**Acknowledgement**

The authors acknowledge the support from National Natural Science Foundation of China (Grant Nos. 11425208, 51232004 and 11372158), and Tsinghua University Initiative Scientific Research Program (No. 2011Z02173).

# Appendix A

According to the paper by Rice et al. (1990), the stress angular functions $\Sigma_{ij}^{I}(\theta,\varepsilon)$ and $\Sigma_{ij}^{II}(\theta,\varepsilon)$ for material 1 (the upper material in Fig. 1) are

$$\Sigma_{rr}^{I}(\theta,\varepsilon) = -\frac{\sinh\left[(\pi-\theta)\varepsilon\right]}{\cosh\pi\varepsilon}\cos\left(\frac{3\theta}{2}\right) + \frac{e^{-(\pi-\theta)\varepsilon}}{\cosh\pi\varepsilon}\cos\left(\frac{\theta}{2}\right)\left(1+\sin^{2}\frac{\theta}{2}+\varepsilon\sin\theta\right) \quad (A1)$$

$$\Sigma_{\theta\theta}^{I}(\theta,\varepsilon) = \frac{\sinh\left[(\pi-\theta)\varepsilon\right]}{\cosh\pi\varepsilon}\cos\left(\frac{3\theta}{2}\right) + \frac{e^{-(\pi-\theta)\varepsilon}}{\cosh\pi\varepsilon}\cos\left(\frac{\theta}{2}\right)\left(\cos^{2}\frac{\theta}{2}-\varepsilon\sin\theta\right) \quad (A2)$$

$$\Sigma_{r\theta}^{I}(\theta,\varepsilon) = \frac{\sinh\left[(\pi-\theta)\varepsilon\right]}{\cosh\pi\varepsilon}\sin\left(\frac{3\theta}{2}\right) + \frac{e^{-(\pi-\theta)\varepsilon}}{\cosh\pi\varepsilon}\sin\left(\frac{\theta}{2}\right)\left[\cos^{2}\left(\frac{\theta}{2}\right)-\varepsilon\sin\theta\right] \quad (A3)$$

$$\Sigma_{rr}^{II}(\theta,\varepsilon) = \frac{\cosh\left[(\pi-\theta)\varepsilon\right]}{\cosh\pi\varepsilon}\sin\left(\frac{3\theta}{2}\right) - \frac{e^{-(\pi-\theta)\varepsilon}}{\cosh\pi\varepsilon}\sin\left(\frac{\theta}{2}\right)\left(1+\cos^{2}\frac{\theta}{2}-\varepsilon\sin\theta\right) \quad (A4)$$

$$\Sigma_{\theta\theta}^{II}(\theta,\varepsilon) = -\frac{\cosh\left[(\pi-\theta)\varepsilon\right]}{\cosh\pi\varepsilon}\sin\left(\frac{3\theta}{2}\right) - \frac{e^{-(\pi-\theta)\varepsilon}}{\cosh\pi\varepsilon}\sin\left(\frac{\theta}{2}\right)\left(\sin^{2}\frac{\theta}{2}+\varepsilon\sin\theta\right) \quad (A5)$$

$$\Sigma_{r\theta}^{II}(\theta,\varepsilon) = \frac{\cosh\left[(\pi-\theta)\varepsilon\right]}{\cosh\pi\varepsilon}\cos\left(\frac{3\theta}{2}\right) + \frac{e^{-(\pi-\theta)\varepsilon}}{\cosh\pi\varepsilon}\cos\left(\frac{\theta}{2}\right)\left[\sin^{2}\left(\frac{\theta}{2}\right)+\varepsilon\sin\theta\right] \quad (A6)$$

For material 2, simply change $\pi$ to $-\pi$ everywhere.

# Appendix B

To investigate the oscillatory period of the stress field, we need compute the stresses at different distance to the crack tip. According to theoretical solution Eq. (1), if two different circle contours with radiuses $r_1$ and $r_2$ satisfy

$$\frac{r_1}{r_2} = e^{\frac{2\pi n}{\varepsilon}}, \qquad (B1)$$

where $n$ is an arbitrary integer, the corresponding normalized stress angular distributions are identical. In our numerical example, the parameters for material 1 and material 2 are $E_1 = 10^6 \, Pa$, $E_2 = 10^5 \, Pa$, $v_1 = v_2 = 0.2$, which lead to $\varepsilon = -0.101$ from Eq. (2)错误!未找到引用源。. Based on Eq. (B1), a normalized stress angular distribution at $r_1$ will reappear at $r_2 = 9.0 \times 10^{-28} r_1 \, (for \, n=1)$ when approaching the crack tip. Hence, it is impossible to precisely capture the oscillatory stress distribution with traditional FEM simulation approaches. A recursive computing scheme, schematically shown in Fig. 4, is adopted instead. The first FEM simulation model is a circle plate with an interface crack on single side under angularly distributed load. For each simulation, we extract the displacement distribution on the circle with the half radius, and apply it on the boundary of next simulation model. Repeating this process, a stress field of any circle can be obtained.

Figure B1 shows the variation of $\sigma_{\theta\theta}\sqrt{r}$ with respect to the distance from the crack tip along the interface. The oscillatory feature of stress field around an interface crack is clear shown. It is found that when approaching the crack tip, the period predicted from simulation is $r_1/r_2 = 9.3 \times 10^{-28}$, which is in good agreement of theoretical solution ($9.0 \times 10^{-28}$). The angular distributions of stresses obtained by the finite element analysis also agree well with the theoretical solution as shown in Fig. B2.

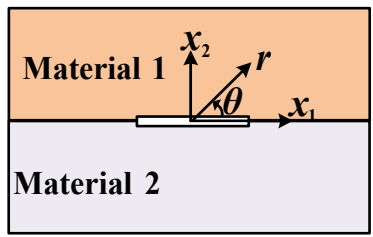

Fig 1 Schematic of an interface crack between two different linear elastic isotropic materials.

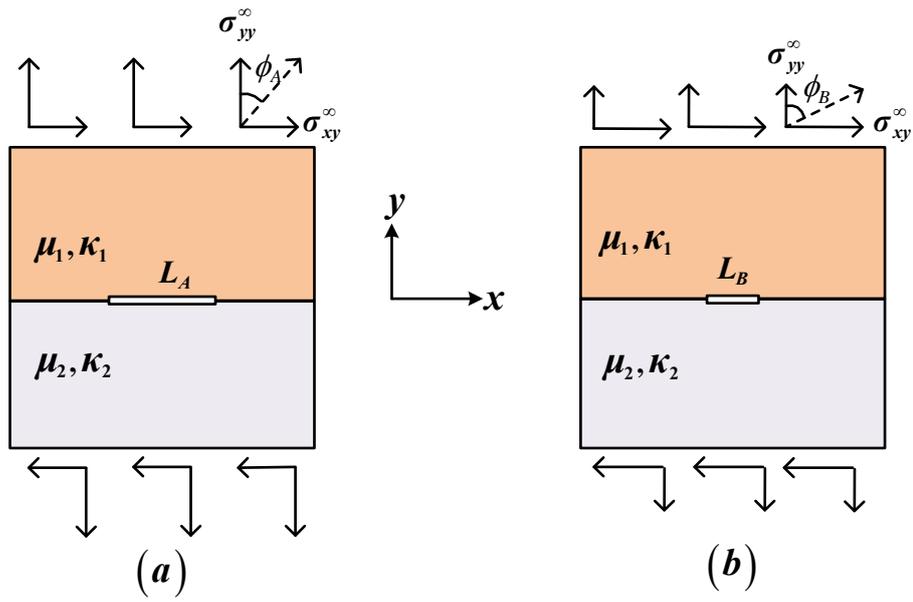

Fig.2 Schematics of two infinite bodies with interface cracks of different lengths ($L_A \neq L_B$) subject to remote loading $\sigma_{yy}^\infty$ and $\sigma_{xy}^\infty$.

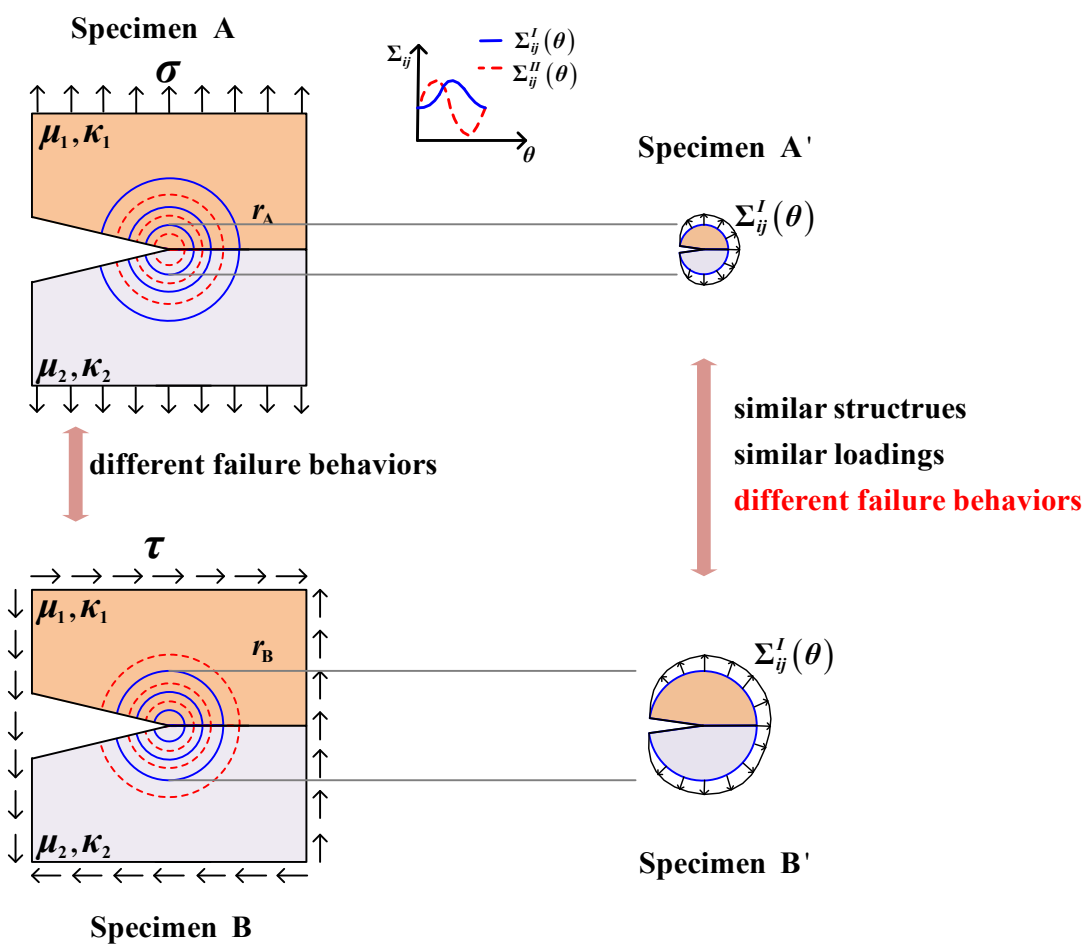

Fig. 3 Schematics of two identical specimens with interface cracks subject to different loadings and their subsystems.

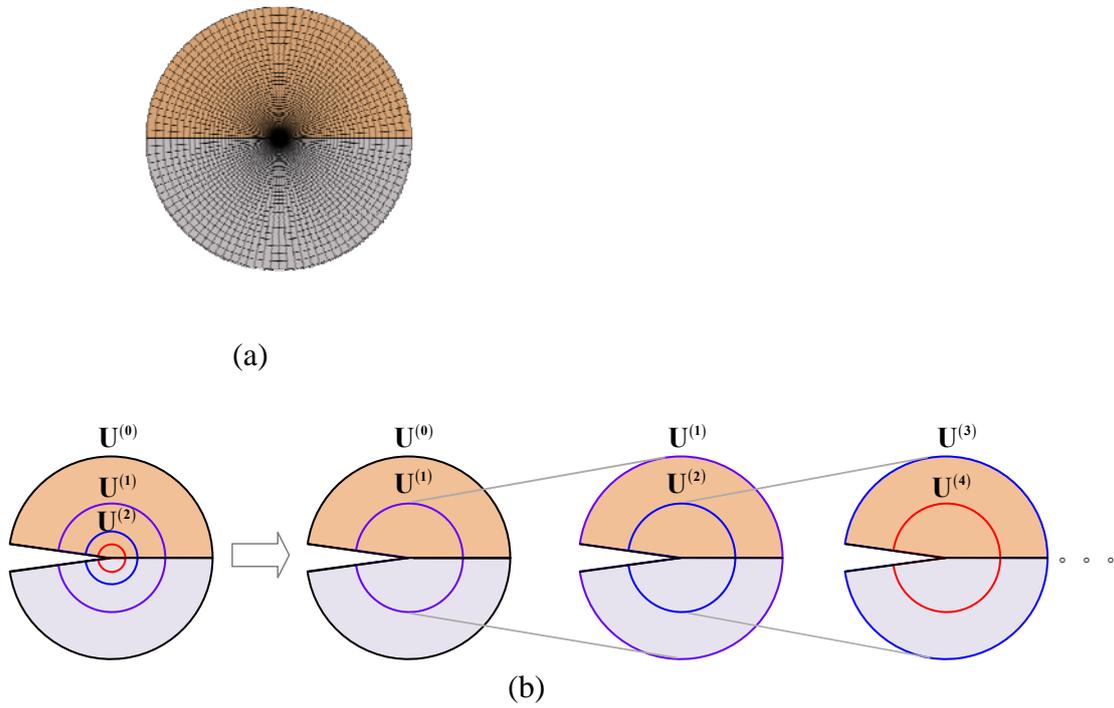

Fig. 4　Simulation models (a) finite element mesh (b) a recursive computing scheme.

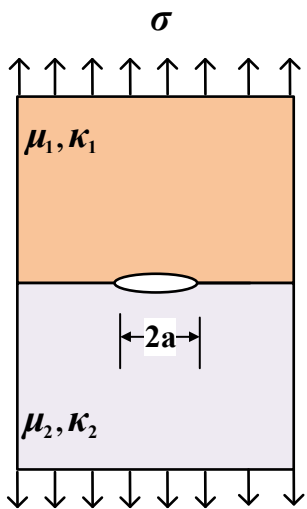

Fig. 5 Schematic of an infinite bi-material body including an interface crack subject to remote uniaxial tension

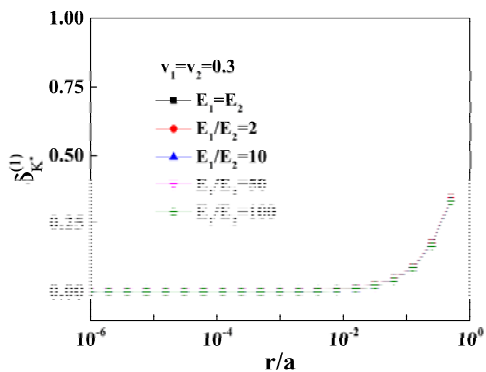
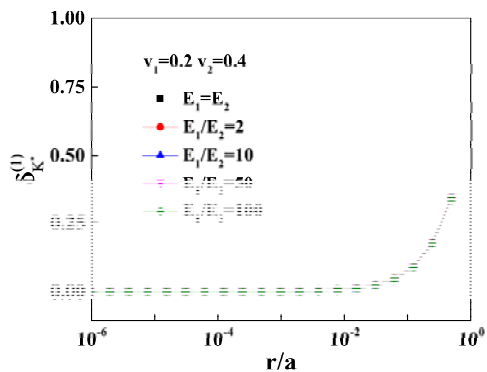

(a)            (b)

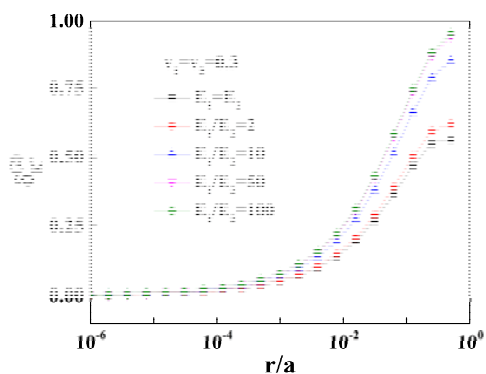
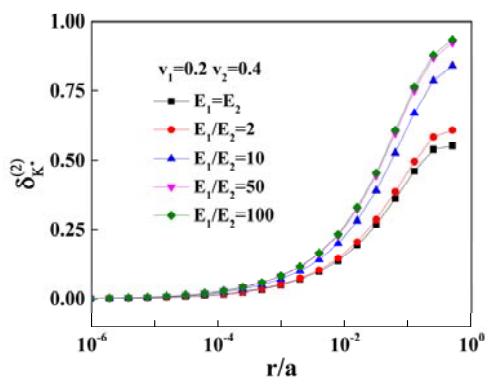

(c)            (d)

Fig.6 The relative errors between the full-field solution and the asymptotic solution for the cases with different stiffness ratios.

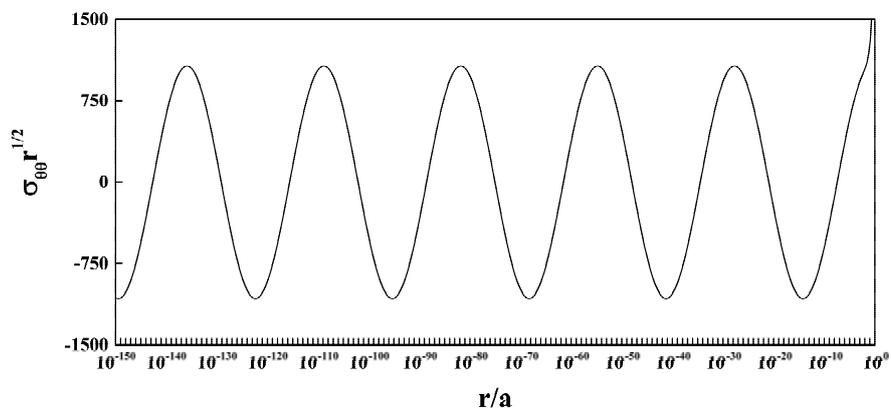

Fig.B1 Variation of the normalized hoop stress along the interface as a function of the normalized distance to the crack tip from simulations.

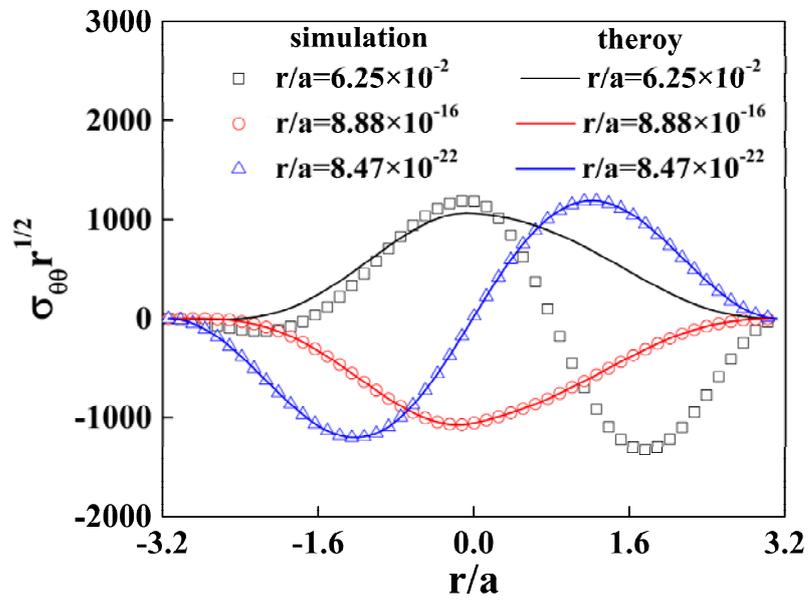

Fig. B2 Comparison of the stress angular distributions between simulations and theoretical predictions